\documentclass[runningheads]{llncs}

\usepackage[pagebackref,breaklinks,colorlinks,allcolors=blue]{hyperref}
\usepackage{graphicx}
\usepackage{amsmath}
\usepackage{amssymb}
\usepackage{booktabs}
\usepackage{caption}
\usepackage{float}
\usepackage{placeins}
\usepackage{microtype}
\usepackage{makecell}

\title{\textbf{Disc-Centric Contrastive Learning for Lumbar Spine Severity Grading}}
\titlerunning{Disc-Centric Contrastive Learning for Lumbar Spine}

\author{
Sajjan Acharya*, Pralisha Kansakar*
}

\authorrunning{S. Acharya \and P. Kansakar}

\institute{
Independent Researchers\\
\email{\{sajjanacharya00, pralishakansakar\}@gmail.com}
}

\begin{document}

\maketitle

\begin{center}
\small * These authors contributed equally to this work.
\end{center}

\begin{abstract}
This work examines a disc-centric approach for automated severity grading of lumbar spinal stenosis from sagittal T2-weighted MRI. The method combines contrastive pretraining with disc-level fine-tuning, using a single anatomically localized region of interest per intervertebral disc. Contrastive learning is employed to help the model focus on meaningful disc features and reduce sensitivity to irrelevant differences in image appearance. The framework includes an auxiliary regression task for disc localization and applies weighted focal loss to address class imbalance. Experiments demonstrate a 78.1\% balanced accuracy and a reduced severe-to-normal misclassification rate of 2.13\% compared with supervised training from scratch. Detecting discs with moderate severity can still be challenging, but focusing on disc-level features provides a practical way to assess the lumbar spinal stenosis.
\end{abstract}

\section{Introduction}
\subsection{Clinical Background}
Recent advances in artificial intelligence (AI) have led to widespread use of data-driven methods in medical imaging, enabling automated analysis at a scale which was previously unattainable. These advanced imaging techniques provide essential data that significantly help specialists to make informed clinical decisions. With recent technological advances, AI has increasingly been applied in healthcare to enhance diagnostic accuracy and operational efficiency. Pinto-Coelho~\cite{pinto2023artificial} highlights how AI-driven image analysis accelerates the interpretation of complex datasets, leading to earlier disease detection and better patient outcomes. Similarly, Khalifa and Albadawy~\cite{khalifa2024ai} emphasize AI's role in revolutionizing clinical workflows, augmenting human expertise by prioritizing urgent cases and handling repetitive tasks.

Lower back pain and lumbar spinal stenosis pose major challenges to healthcare systems worldwide.~\cite{gbd2023global}. Magnetic Resonance Imaging (MRI) serves as the primary diagnostic modality for identifying underlying spinal pathologies, offering superior soft-tissue contrast essential for surgical planning~\cite{hallinan2021deep}.

Lumbar Spinal Stenosis(LSS) is characterized by a narrowing of the spinal canal that compresses neural and vascular structures~\cite{munakomi2024lumbar}. The condition results from spinal degeneration: discs lose height, load on the spine changes, ligaments thicken, and bone spurs form, which can narrow the spinal canal.

Clinically, the hallmark of LSS is neurogenic claudication, characterized by pain, paresthesia, or cramping in the lower back and legs that worsens with lumbar extension and improves with flexion~\cite{lurie2016management}. Diagnosis relies heavily on MRI for visualizing soft tissue structures and grading central, lateral, and foraminal stenosis.

Radiological assessment of stenosis severity is inherently disc-centric and volumetric. For each intervertebral level, clinicians synthesize visual evidence across multiple contiguous MRI slices to assign a single diagnostic grade~\cite{hallinan2021deep}. This process is cognitively complex and subjective, requiring high-level subspecialist expertise~\cite{lurie2016management}.

However, rising imaging volumes create unprecedented workloads for radiologists, increasing the risk of diagnostic variability and delays in patient care. Automated deep learning models offer a promising avenue to provide consistent, rapid, and objective grading of spinal pathologies~\cite{khalifa2024ai}.

\subsection{Challenges in Automated MRI-Based Severity Classification}
Despite numerous contributions and successes in medical imaging, developing robust deep learning models for spinal analysis remains challenging. The high dimensionality of 3D MRI volumes creates a significant bottleneck for deep learning models. A standard high-resolution MRI series contains millions of voxels, making full-volume processing computationally expensive, memory-intensive and impractical for many deep learning architectures~\cite{cciccek20163d}. 

Pathological conditions often exhibit extreme class imbalance, where ``Normal'' cases vastly outnumber ``Severe'' pathologies, biasing models toward majority classes. This imbalance biases models toward majority classes and degrades sensitivity to clinically critical cases~\cite{johnson2019survey}. Importantly, not all classification errors carry equal clinical risk. Misclassifying a severe pathology as normal can lead to delayed diagnosis and adverse outcomes, a distinction that is not adequately captured by conventional metrics such as overall accuracy or macro-F1 score~\cite{saito2015precision}.

\subsection{Our Core Hypothesis}
We hypothesize that the primary bottleneck in automated lumbar stenosis grading lies in learning disc-level representations that are robust to non-diagnostic appearance variations, rather than in the choice of downstream classifier. Specifically, if an encoder learns disc-consistent features within a localized anatomical ROI, downstream classification becomes simpler, safer, and more robust. Our approach focuses on promoting such anatomically consistent disc-level representations to improve grading performance across severity levels.

\subsection{Contributions}
We investigate a disc-centric contrastive learning framework for lumbar spine grading that leverages a single, anatomically localized region of interest (ROI) for each intervertebral disc, derived from sparse coordinate annotations. Contrastive pretraining is conducted by applying stochastic augmentations to this disc-level view, thereby promoting invariance to non-diagnostic variations in appearance while retaining clinically salient morphological features. The resulting representations are assessed via linear probing and partial fine-tuning protocols. We introduce safety-oriented evaluation metrics that explicitly quantify severe-to-normal misclassification rates, complementing standard accuracy-based measures. Our results demonstrate that contrastive pretraining, rather than increased architectural complexity, plays a central role of performance improvements in low-supervision areas.

\section{Related Work}
\subsection{Spine MRI Analysis}
Early spine MRI analysis relied on handcrafted features and heuristic rules, but the field has since shifted toward deep learning. Convolutional Neural Network (CNN) methods have been widely applied to vertebral labeling, disc degeneration grading, and stenosis detection. For instance, Hallinan et al.~\cite{hallinan2021deep} utilized a fully supervised, two-stage CNN pipeline that relies on dense radiologist annotations to detect regions of interest (ROIs) and classify stenosis. Similarly, recent works like MRI Spine Pathology Detection by Subramanian, B., et al.~\cite{subramanian2025ai} demonstrate the efficacy of Vision Transformers and MedSAM on massive proprietary datasets (2M+ scans), such approaches are often computationally demanding and rely on large-scale proprietary datasets. 

\subsection{Volumetric Models and Reconstruction}
To complement slice-level analysis, reconstruction-based self-supervised approaches seek to learn rich volumetric representations by recovering missing image regions. These methods emphasize detailed image reconstruction, contributing valuable structural information that can support downstream pathology grading.

Concurrently, deep learning has begun to improve upstream acquisition. Zerunian et al.~\cite{zerunian2024fast} showed that deep learning reconstruction can significantly reduce acquisition times and improve image quality. Similarly, Hammernik et al.~\cite{hammernik2018learning} showed that learning-based variational networks enable high-quality MRI reconstruction by effectively modeling image priors within an end-to-end framework. While these methods capture structural information useful for downstream tasks, they are computationally expensive and not optimized for disc-level severity grading. Our framework instead leverages slice-level embeddings with coordinate guidance, providing clinically relevant, efficient representations.

\subsection{Anatomical Localization}
Precise localization is a prerequisite for automated diagnosis. Various approaches for anatomical localization have been used, for instance, Sayed et al.~\cite{sayed2025automatic} recently demonstrated the utility of YOLOv8 for detecting intervertebral discs, achieving high precision by treating spinal structures as discrete objects.

Regression-based localization offers a streamlined alternative for vertebral detection and alignment, reducing spatial variability. Genç et al.~\cite{gencc2025deep} employ a standard supervised learning approach, where deep neural networks are trained to map MRI images directly to clinical grades using extensive labeled data.

\subsection{Contrastive Learning in Medical Imaging}
Contrastive learning has been applied to medical imaging for self-supervised representation learning, particularly in radiology and histopathology. Prior works focus on instance-level or patient-level contrastive objectives. Disc-level contrastive learning for spine MRI, particularly under weak, coordinate-based supervision, remains an area with significant potential for exploration.

Beyond diagnostic tasks, self-supervised learning has shown significant promise in upstream MRI tasks. Ekanayake et al.~\cite{ekanayake2025cl} demonstrated that contrastive learning could effectively regularize undersampled MRI reconstruction, allowing models to learn robust structural features without full supervision. Building on this success, our framework extends contrastive principles to the downstream task of severity grading, leveraging coordinate guidance to define clinically relevant positive pairs.

In the domain of semi-supervised segmentation, Chaitanya et al.~\cite{chaitanya2020contrastive}  demonstrated that exploiting the structural similarity of corresponding slices across different patient volumes can significantly improve contrastive representation learning. While their approach assumes rough implicit alignment to define positive pairs, our coordinate-guided framework explicitly enforces anatomical correspondence using weak supervision, allowing us to generate highly precise positive pairs for pathology grading even in unaligned or heterogeneous datasets.

While recent multimodal approaches like LumbarCLIP~\cite{le2025revolutionizing} have successfully utilized radiology reports to guide contrastive learning for binary lower back pain diagnosis, they rely on the availability of paired expert text, which is often unstructured or unavailable in public datasets. In contrast, our coordinate-guided framework leverages sparse geometric annotations, universally available and cheap to acquire, to learn fine-grained, level-specific representations suitable for severity grading rather than just binary classification. 

\section{Methodology}

\subsection{Dataset and Preprocessing}

\subsubsection{Data Background}
We utilized the RSNA 2024 Lumbar Spine Degenerative Classification Dataset~\cite{richards2025rsna} consisting of MRI scans of 2,697 patients with a total of 8,593 image series from 8 institutions across 6 countries and 5 continents. 

\subsubsection{Data Composition}
Each study in the dataset typically includes multiple MRI sequences. For this work, we focus exclusively on the Sagittal T2-weighted scans (T2), which provide optimal contrast for assessing intervertebral disc height, hydration status, and spinal canal stenosis.

\subsubsection{Clinical Annotation}
The dataset provides annotations for lumbar spinal canal stenosis across five intervertebral disc levels (L1/L2 to L5/S1). The severity of each label is graded on a three-point ordinal scale:
\begin{itemize}
    \renewcommand{\labelitemi}{\textbullet}
    \item \textbf{Normal/Mild:} No or insignificant narrowing.
    \item \textbf{Moderate:} Significant narrowing but without critical compression.
    \item \textbf{Severe:} Critical narrowing often requiring surgical intervention.
\end{itemize}

Additionally, the dataset also comprises the coordinates of the pathologies that are annotated for each of the patients. These coordinates pinpoint the exact location of the pathologies, providing ground truth for both localization and classification tasks.

\subsubsection{DICOM Standardization}
The raw dataset consists of MRI series stored in the DICOM standard. To facilitate efficient training with standard deep learning libraries, we implement a conversion pipeline to transform high-dynamic-range DICOM files into standardized PNG format. The process implements the following steps:
\begin{itemize}
    \renewcommand{\labelitemi}{\textbullet}
    \item \textbf{Pixel Intensity Normalization:} The raw pixel array is extracted from the DICOM object and normalized to a floating-point range. 
    \item \textbf{Padding and Coordinate-Guided Cropping:} To preserve anatomical context for pathologies located near the image boundaries, the normalized slices are first padded with constant values. Subsequently, utilizing the provided spatial coordinates $(x, y)$, a fixed-size Region of Interest (ROI) is extracted centered specifically on the indicated spinal level. 
    \item \textbf{Format Conversion:} The extracted patches are rescaled to an 8-bit integer range $[0, 255]$ and saved as lossless PNG images. This conversion significantly reduces storage overhead and streamlines data loading throughput during the training phase.
\end{itemize}


\subsubsection{Disc-Centric Data Partitioning}
All dataset splits are performed at the intervertebral disc level. This ensures that slices corresponding to the same disc do not appear across different splits. Each disc is uniquely identified by the patient ID, scan type, and anatomical disc level, and is treated as a distinct sample for training and evaluation.
Stratified splitting based on severity grades is applied to preserve the label distribution across the training, validation, and test sets.

\subsection{ROI Localization via Coordinate Regression}
To automate the identification of relevant anatomical structures in unseen data via cropping, we employ a localization module designed to regress spatial coordinates. The regression module predicts intervertebral disc center coordinates $(x, y)$ from MRI slices as an auxiliary task to the primary severity classification, which is trained and evaluated independently from the primary classification.

The main classification task operates on individual 2D slices, which serve as the primary input to the network. For the auxiliary ROI localization task, three consecutive slices are stacked to form a 2.5D input, providing additional spatial context to improve disc-level coordinate prediction.

\subsection{Encoder Architecture}
We employ a convolutional neural network encoder $E(\cdot)$ based on ResNet-18, adapted for single-channel MRI input. The encoder maps each slice to a 512-dimensional latent representation:
\begin{equation}
    z_i = E(x_i)
\end{equation}

The final fully connected layer is removed to prevent early entanglement of representations with task-specific decision boundaries. This allows the encoder to serve as a general-purpose anatomical feature extractor.

\subsection{Contrastive Representation Learning}

\subsubsection{Contrastive Pair Construction}
We employ a multi-positive instance-level contrastive learning framework to pretrain the encoder on extracted Regions of Interest (ROIs) from sagittal T2-weighted spinal MRI. Since ground-truth coordinates are provided for the single most representative mid-sagittal slice of each disc level, as provided by the dataset annotations, we employ an augmentation-based contrastive strategy. For each unique disc, we generate V=3 stochastic views of its representative slice using augmentations.
This results in a multi-positive contrastive setup, which differs from standard instance discrimination.

\subsubsection{Contrastive Training}
We define a feature encoder $f(\cdot)$ and a projection head $g(\cdot)$ that maps an input slice $x$ to a latent representation $z = g(f(x))$. The goal is to optimize the contrastive network in such a way that the distance between representations in the latent manifold reflects their anatomical relationship.

By maximizing the cosine similarity between augmented views of the same disc-level ROI and minimizing it for views originating from different discs, the model is encouraged to ignore transient noise and acquisition-related artifacts, focusing instead on invariant morphological characteristics relevant to intervertebral disc pathology.

\subsection{Supervised Adaptation and Differential Fine-tuning}
The final stage of our framework involves transitioning from self-supervised representation learning to a supervised classification task. This phase is designed to map the learned manifold to three clinical grades of disc pathology: Normal, Moderate, and Severe.

\subsubsection{Permutation-Invariant Disc Representation}
For supervised severity grading, we employ a permutation-invariant disc representation based on the Deep Sets framework. Given a set of disc embeddings $\{ z_{d,1}, \ldots, z_{d,S} \}$, where each embedding corresponds to a region-of-interest (ROI) instance extracted from the same intervertebral disc, a pooled representation is computed through mean aggregation. This design inherently accommodates inputs of variable size and provides a straightforward extension path for future volumetric analyses. In the present setting, disc-level representations are constructed from a single representative ROI ($S=1$), consistent with the available supervision. The permutation-invariant formulation allows the model to seamlessly generalize to multi-slice inputs ($S > 1$) during future inference without requiring architectural modifications or retraining.

\subsubsection{Differential Fine-tuning}
Rather than training all layers equally, the ResNet-18 backbone is partially frozen to preserve the robust anatomical features learned during the contrastive phase. This prevents “catastrophic forgetting” of the spatial consistency learned in the contrastive stage while balancing stability and adaptability, allowing high-level features to specialize for pathology severity without destroying spatial consistency learned during pretraining. We also employ a differential fine-tuning strategy in which the pretrained backbone is optimized with a lower learning rate than the classification head.

\subsubsection{Focal Loss for Imbalanced Pathology}
The dataset used has severe class imbalance, with Normal cases significantly outnumbering Severe pathologies. Thus, to address this, we replace the standard cross-entropy loss with a weighted focal loss.

\subsection{Losses}

\

\subsubsection{Focal Loss}
Focal loss~\cite{lin2017focal} extends cross-entropy by down-weighting well-classified examples and focusing training on harder, misclassified samples, while class-specific weighting further compensates for imbalanced class frequencies.

For a given sample with ground-truth class $y$ and predicted class probability $p_y$, the weighted focal loss is defined as:
\begin{equation}
    \mathcal{L}_{\text{focal}} = -\alpha_y (1 - p_y)^{\gamma} \log(p_y)
\end{equation}
where:
\begin{itemize}
    \renewcommand{\labelitemi}{\textbullet}
    \item $p_y$ is the predicted probability for the true class $y$,
    \item $\alpha_y$ is a class-specific weighting factor inversely proportional to class frequency,
    \item $\gamma$ is the focusing parameter that controls the strength of hard-example mining.
\end{itemize}

When $\gamma = 0$, the focal loss reduces to weighted cross-entropy. Increasing $\gamma$ places greater emphasis on misclassified and ambiguous samples, which is particularly beneficial in medical imaging settings where severe cases are rare but clinically critical.

\subsubsection{Multi-positive NT-Xent (InfoNCE) Loss}
NT-Xent loss is commonly employed in self-supervised and supervised contrastive learning, especially when an anchor is associated with multiple positive samples. This formulation corresponds to the loss used in Supervised Contrastive Learning (SupCon)~\cite{khosla2020supervised} and its extensions to unsupervised, instance-level, and multi-positive group- or cluster-based settings, such as in SimCLR.

For a given anchor representation $z_i$, let $\mathcal{P}(i)$ denote the set of positive samples corresponding to the same disc. The multi-positive contrastive loss for anchor $i$ is defined as:
\begin{equation}
\mathcal{L}_i = -\frac{1}{|\mathcal{P}(i)|} \sum_{p \in \mathcal{P}(i)}
\log \frac{\exp\big(\mathrm{sim}(z_i, z_p) / \tau \big)}
{\sum_{k \neq i} \exp\big(\mathrm{sim}(z_i, z_k) / \tau\big)}
\end{equation}
where, $\mathrm{sim}(\cdot, \cdot)$ denotes cosine similarity and $\tau$ is a temperature hyperparameter controlling the concentration of the distribution.

Equivalently, this can be expressed in a numerically stable log-sum-exp form as:
\begin{equation}
\mathcal{L}_i = -\frac{1}{|\mathcal{P}(i)|} 
\Bigg[
\log \sum_{p \in \mathcal{P}(i)} \exp \frac{\mathrm{sim}(z_i, z_p)}{\tau} 
-
\log \sum_{k \neq i} \exp \frac{\mathrm{sim}(z_i, z_k)}{\tau} 
\Bigg]
\end{equation}

The overall objective is obtained by averaging $\mathcal{L}_i$ over all anchors in the batch:
\begin{equation}
\mathcal{L} = \frac{1}{N} \sum_{i=1}^{N} \mathcal{L}_i
\end{equation}

When each anchor has a single positive ($|\mathcal{P}(i)| = 1$), this formulation reduces to the standard NT-Xent loss used in SimCLR. Both formulations are mathematically equivalent and provide stable gradients during optimization.

\section{Implementation Details and Experiments}

\subsection{Dataset Preparation and Splitting Strategy}
\begin{figure}[t!] 
    \centering
    \includegraphics[width=\textwidth]{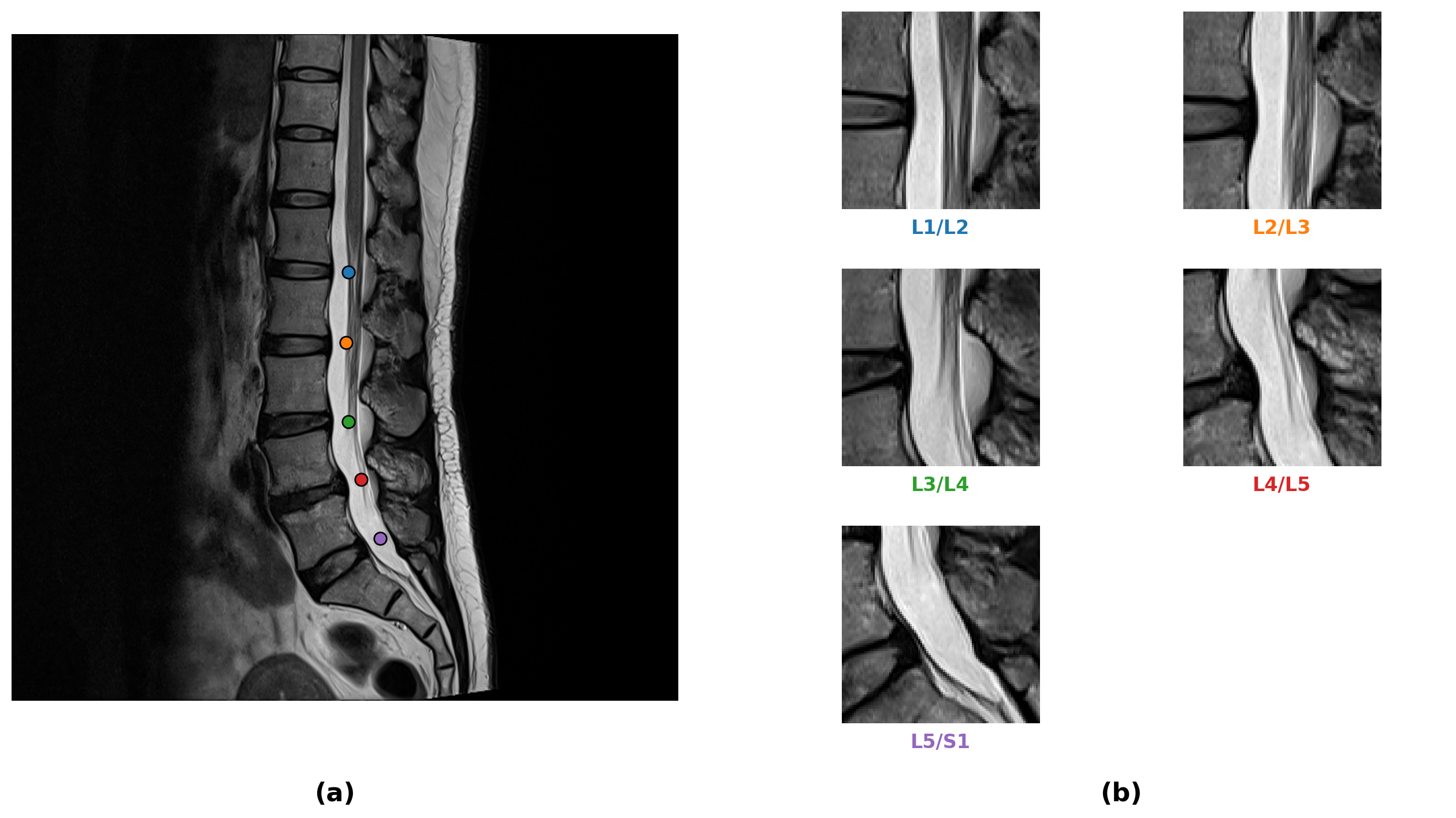} 
    
    \vspace{-10pt} 
    
    \caption[Data Preprocessing Visualization]{Visualization of disc localization and cropping.
    \textbf{(a)}~Sagittal T2-weighted lumbar spine MRI slice showing the five intervertebral disc locations (L1/L2 through L5/S1) marked with colored dots. 
    \textbf{(b)}~Corresponding $96\times96$ pixel cropped Regions of Interest (ROIs) for each disc. Colors of the dots in (a) match the labels in (b).}
    \label{fig:cropped_demo}
    
    \vspace{-15pt} 
\end{figure}

Experiments are conducted on sagittal T2-weighted lumbar spine MRI scans. For each intervertebral disc level, a single representative Region of Interest (ROI) is extracted based on the provided pathology coordinates. We extracted a $96\times96$ pixel crop centered on the disc individually as illustrated in Figure~\ref{fig:cropped_demo}. For pathologies located near image boundaries, constant padding is applied prior to cropping to maintain a consistent ROI size.


To prevent data leakage across experiments, all dataset splits are performed at the disc level, ensuring that images from the same disc of the same study never appeared in both training and validation sets. Stratified splitting is then applied at the disc level based on pathology severity labels (Normal, Moderate, Severe), preserving class distribution across splits. This strategy prevents data leakage; random splitting could inflate performance due to near-identical anatomical content across splits.

All reported metrics are computed on held-out validation discs. To isolate the diagnostic performance of the contrastive encoder from localization errors, all classification metrics are computed using ground-truth disc coordinates. The regression-based localization module is evaluated separately to demonstrate the feasibility of an automated pipeline under minimal supervision. For 2.5D experiments, three adjacent slices are stacked channel-wise to encode limited volumetric context.

\subsection{Image Preprocessing and Augmentation}
All MRI slices are resized to $224 \times 224$ pixels and normalized using channel-wise statistics consistent with grayscale medical imaging. For contrastive pretraining, augmentations are applied to encourage invariance to non-diagnostic variations in the images. Augmentations include:
\begin{itemize}
    \renewcommand{\labelitemi}{\textbullet}
    \item Random spatial cropping and resizing (scale 0.8 - 1.0)
    \item Random horizontal flipping
    \item Random rotation ($\pm 15^\circ$)
    \item Intensity jittering (brightness, contrast, and saturation)
\end{itemize}

These transformations preserve disc morphology while simulating acquisition variability, minor patient motion, and scanner-induced intensity shifts. For supervised fine-tuning and evaluation, augmentations are disabled, and only deterministic preprocessing (resizing and normalization) is applied to ensure consistent inference behavior.

\subsection{Contrastive Pretraining Configuration}
Contrastive pretraining is performed using the encoder-projection architecture. The encoder is a ResNet-18 initialized from random weights, while the projection head consists of two fully connected layers with batch normalization and ReLU activation. 

Each input slice generates V augmented views, forming a multi-positive contrastive group. 
The network is optimized using the multi-positive NT-Xent loss, with a temperature parameter $\tau = 0.1$. Gradient norm clipping is applied to stabilize training under low-temperature settings. The training was done for 60 epochs utilizing a \texttt{CosineAnnealingLR} scheduler \cite{loshchilov2016sgdr}.

\subsection{Supervised Fine-tuning}
Following pretraining, the model was adapted for disc-level pathology classification. We applied a split learning rate strategy. The backbone (layer4) was optimized at a rate of $5 \times 10^{-5}$, while the classification head was optimized at a higher rate of $5 \times 10^{-4}$. All earlier layers (layers 1--3) remained frozen to preserve low-level anatomical features.

To mitigate the significant class imbalance inherent in the dataset, training was guided by a Weighted Focal Loss with a focusing parameter $\gamma = 2.0$ and class-specific weights $\alpha = [0.8, 4.0, 5.0]$ for the Normal/Mild, Moderate, and Severe classes, respectively. The fine-tuning process ran for 40 epochs with a batch size of 24 discs, utilizing a \texttt{ReduceLROnPlateau} scheduler \cite{moreira1995neural} (patience=4, factor=0.5) to dynamically adjust the learning rate based on validation balanced accuracy.

\subsection{Regression-Based Localization Module (Auxiliary Task)}
The ROI localization network is built on a ResNet18 backbone pretrained on \texttt{IMAGENET1K\_V1}, modified to accept three consecutive slices stacked as a 2.5D input of size $256\times256$ pixels. Disc-level identifiers are embedded into a 32-dimensional vector and concatenated with backbone features. 

The inputs are processed through a 256-unit fully connected layer with 0.5 dropout, followed by a regression head predicting normalized 2D coordinates for the disc of interest. Only the regression branch is active for this auxiliary task. The model was trained with a batch size of 32 using the AdamW optimizer (learning rate $1\times10^{-4}$, weight decay 0.01) and a \texttt{ReduceLROnPlateau} scheduler (patience 3, factor 0.1). The regression module was optimized using Smooth L1 loss, and performance during validation was assessed using the root mean squared error (RMSE).

While not our primary contribution, this module enables a fully automated inference pipeline, allowing the classification model to be applied to unannotated MRI scans. The localization component thus strengthens the system’s practical applicability and scalability.

\subsection{Hardware and Software}
All models were implemented in PyTorch and trained on an NVIDIA RTX-3060 GPU. Convergence was monitored using the validation loss for contrastive training and balanced accuracy for the fine-tuned classification models. The best-performing weights were selected for further downstream tasks and application.

\section{Results}

\subsection{Contrastive Training Results}

\begin{figure}[htbp]
    \centering
    \includegraphics[width=\linewidth]{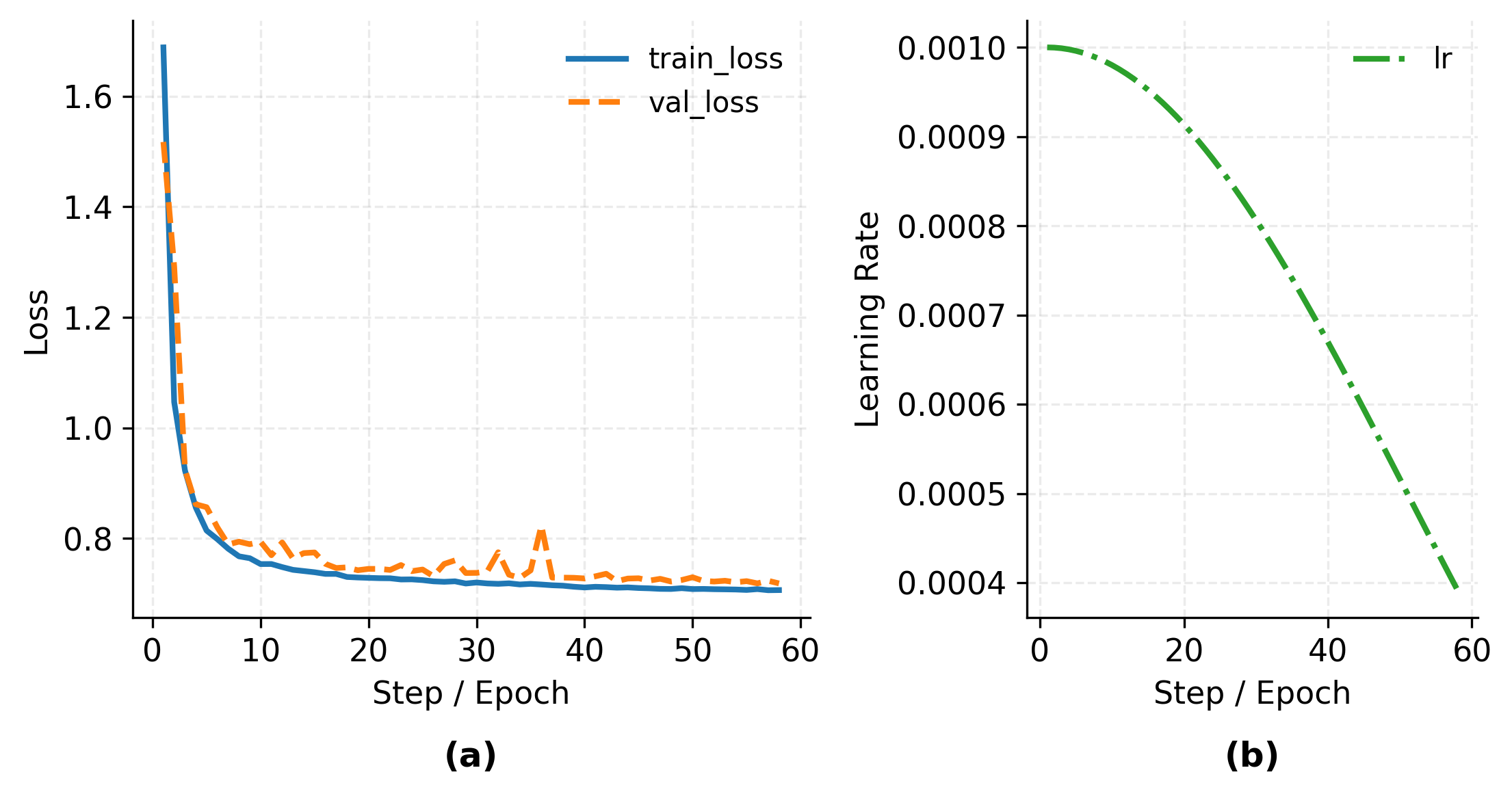}
    \caption{Training dynamics of the contrastive representation learning framework. 
    (a) Evolution of the multi-positive InfoNCE loss for training (blue) and validation (orange) sets over 60 epochs. 
    (b) Learning rate trajectory utilizing a Cosine Annealing schedule.}
    \label{fig:contrastive_training}
\end{figure}

The primary objective was to map high-dimensional MRI cropped disc images into a latent space where anatomical structure is preserved despite augmentation-induced variability. The convergence of the loss to a stable baseline (~0.7) indicates that the model has successfully learned augmentation-invariant representations (Figure~\ref{fig:contrastive_training}). Gradient Norm Clipping proved useful when training with a low temperature parameter in the InfoNCE Loss, as it mitigated instability and prevented late-stage oscillations.

\subsection{Disc Level Fine-tuned Classifier}

\begin{figure}[htbp!]
    \centering
    \includegraphics[width=\linewidth]{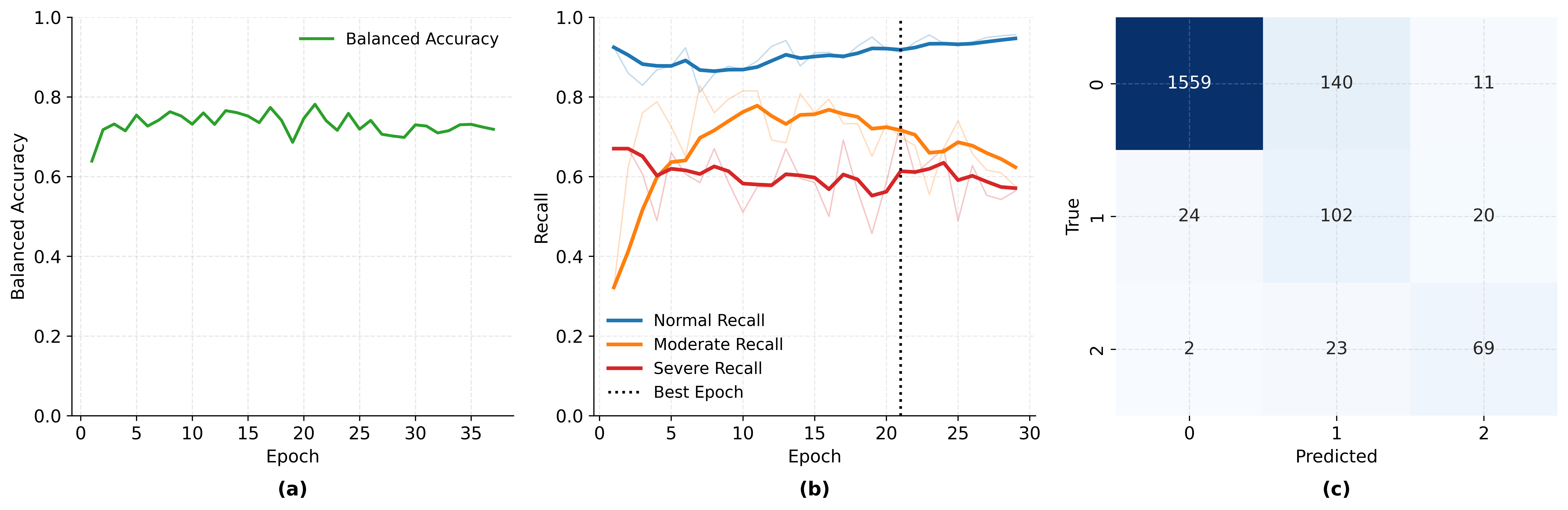}
    \caption{Clinical performance and convergence analysis. 
    (a) Validation balanced accuracy over 37 epochs. 
    (b) Class-specific recall trajectories (Normal, Moderate, Severe); bold lines represent smoothed trends (EMA), dashed line at Epoch 21 marks the optimal checkpoint. 
    (c) Confusion matrix at Epoch 21.}
    \label{fig:disc_classifier}
\end{figure}

Contrastive pre-training enabled the model to learn discriminative anatomical features, achieving a balanced accuracy of 78\%. As illustrated in Figure~\ref{fig:disc_classifier}, the model organizes normal anatomy and progressive stages of degeneration in a structured latent space, instead of overfitting to label frequencies. This is evidenced by consistently high recall for the Normal class and a gradual, progressive improvement in recall across pathological classes, indicating that the learned representations reflect meaningful morphological differences.

The application of weighted focal loss effectively mitigated class imbalance by prioritizing the Severe class, ensuring that clinically critical cases had substantial influence throughout training. Analysis of misclassifications indicates the model's reliability in a clinical context: the majority of errors occur between adjacent severity grades, consistent with their visual similarity. Misclassification of Severe cases as Normal was rare, occurring in only 2.1\% of instances (2/94). The checkpoint from epoch 21 was selected as the final model, providing the optimal trade-off between stable recall across pathological classes and overall performance. Although later epochs showed modest gains in precision, recall for pathological findings became less consistent, making epoch 21 the most robust choice for potential clinical screening use.

\subsection{ROI Metrics}
\begin{figure}[htbp!]
    \centering
    \includegraphics[width=0.8\linewidth]{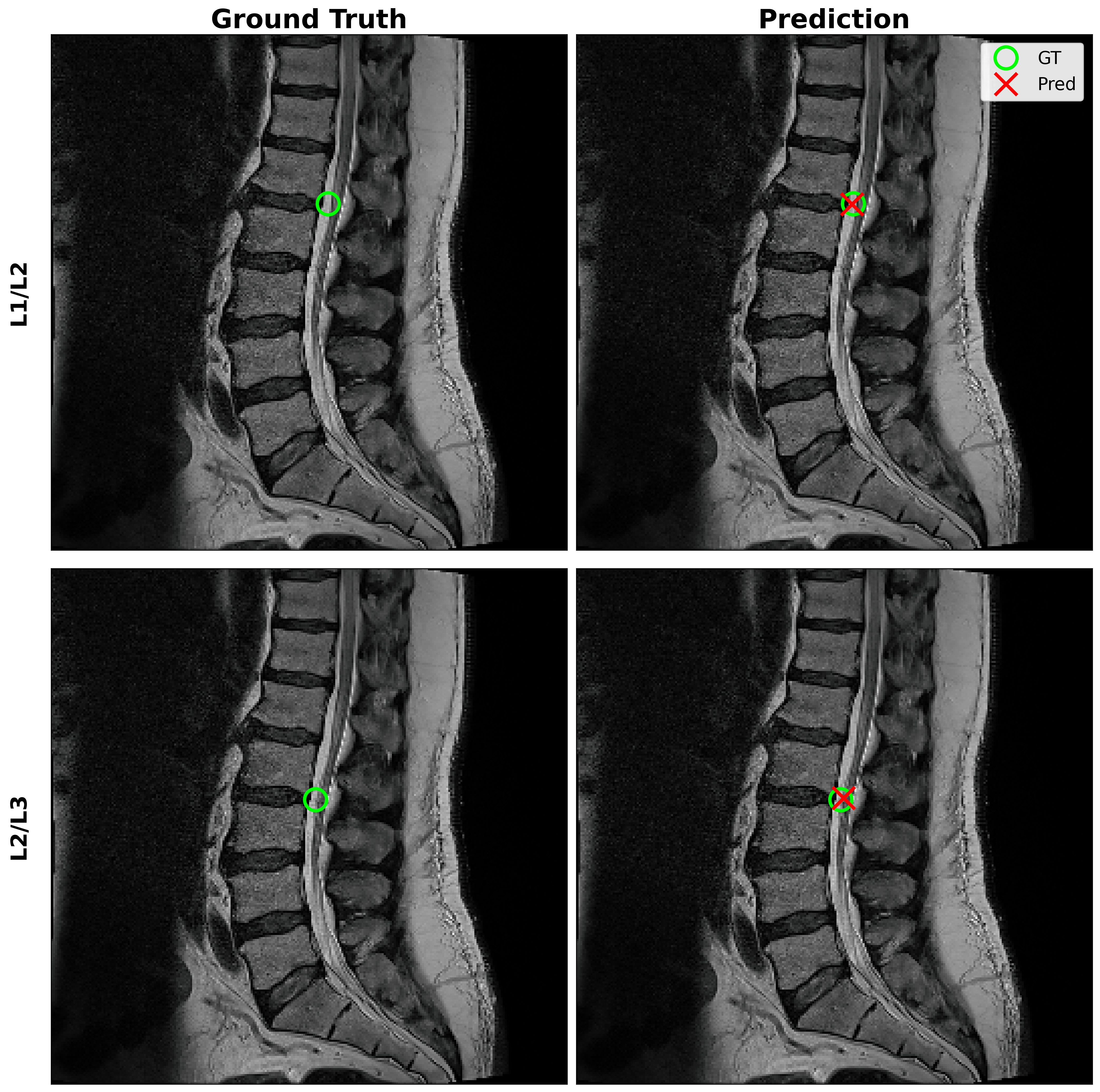}
    \caption{Visualization of model performance on a representative patient series (L1/L2, L2/L3). Each row corresponds to a vertebral level. Left: Original MRI with Ground Truth (green circle). Right: Model prediction (red cross) versus Ground Truth. }
\end{figure}

The model was trained for 40 epochs, achieving its best performance at epoch 31, with an overall loss of 0.0016 and a RMSE of 10.18 pixels on the test dataset. These results demonstrate that the proposed ROI network can reliably predict anatomical coordinates of intervertebral discs with high precision and stable quantitative performance. This serves as an effective auxiliary tool to support our classification task.

\FloatBarrier
\section{Metrics and Discussion}
We compare three configurations to evaluate the contribution of contrastive pretraining using an identical ResNet-18 backbone Table~\ref{tab:results_table}. First, we train the network from scratch with random initialization as a supervised baseline, representing performance without self-supervised priors. Second, we evaluate a Linear Probe, where the contrastive encoder is frozen and only the final classification layer is optimized. This evaluates how well the frozen encoder features capture anatomical structure relevant for classification. Finally, we report our proposed Fine-tuned Disc-Aware ResNet, where encoder weights are initialized from contrastive pretraining and then fine-tuned for severity grading.

\begingroup
\setlength{\abovecaptionskip}{4pt}  
\setlength{\belowcaptionskip}{4pt}  
\setlength{\intextsep}{6pt}         
\setlength{\tabcolsep}{4pt}         
\begin{table}[H]

\centering
\caption{Disc-level classification performance of different models. Columns: Recall for Normal, Moderate, Severe discs, Overall Accuracy, Balanced Accuracy, and Severe-to-Normal misclassification rate. Best values per column are bolded.}
\label{tab:results_table}
\begin{tabular}{l c c c c c c}
\hline
\textbf{Model} & \textbf{R(N)} & \textbf{R(M)} & \textbf{R(S)} & \textbf{Acc} & \textbf{Bal Acc} & \textbf{S-N Err} \\
\hline
\makecell[l]{ResNet + Classifier \\ (from scratch)} 
& 0.812 & 0.377 & 0.755 & 0.776 & 0.648 & 5.32\% \\
\makecell[l]{Linear Probe \\ (Contrastive Encoder)} 
& 0.875 & \textbf{0.719} & 0.564 & 0.848 & 0.719 & 5.32\% \\
\makecell[l]{Fine-tuned Disc-Aware \\ ResNet (Ours)} 
& \textbf{0.912} & 0.699 & \textbf{0.734} & \textbf{0.887} & \textbf{0.781} & \textbf{2.13\%} \\
\hline
\end{tabular}
\end{table}
\endgroup

\subsubsection{Impact of Contrastive Pretraining and Fine-tuning}
The results presented in Table~\ref{tab:results_table} demonstrates that contrastive pretraining provides a clear initialization advantage over standard supervised learning. The linear probe (frozen contrastive encoder) significantly outperforms the scratch-trained ResNet-18 baseline (Balanced Accuracy: 0.719 vs. 0.648), suggesting that the encoder captures robust anatomical structure even in the absence of label supervision. However, limited recall for Severe cases (56.4\%) by linear probe shows that frozen features emphasize anatomical structure but are insufficient for capturing severity-specific pathology. The Fine-tuned Disc-Aware ResNet achieves the best overall performance (Balanced Accuracy: 0.781), demonstrating that contrastive pretraining establishes a strong feature space, but partial fine-tuning is required to adapt these representations to clinical severity grading.

\subsubsection{Clinical Safety and Error Analysis}
A critical advantage of our proposed method is the reduction of clinically dangerous errors. The fine-tuned model reduces Severe-to-Normal misclassifications from 5.3\% (baseline) to 2.1\%. This improvement is particularly important in a screening setting, where missed severe cases carry high clinical risk. Across all models, performance on the Moderate class remains limited, with lower recall than both Normal and Severe cases. This likely reflects the intrinsic ambiguity of intermediate degeneration stages and suggests that future work may benefit from ordinal regression or uncertainty-aware loss functions to better handle these intermediate or borderline cases.

\section{Conclusion}
This study demonstrates that anatomically consistent, disc-level representation learning can substantially improve the safety and stability of lumbar spinal stenosis grading, even in the absence of volumetric modeling or large-scale supervision. While moderate-grade pathology remains difficult to classify due to inherent visual ambiguity, the proposed framework provides a stable and scalable basis for automated severity assessment and reduces clinically critical misclassifications. Overall, this study supports disc-centric contrastive representation learning as a clinically aligned approach for musculoskeletal MRI analysis.

\section{Challenges and Future Enhancements}
While the proposed framework achieves strong overall performance, accurately grading moderate disease severity remains challenging, as the visual differences between neighboring grades are often subtle. The current focus on sagittal T2-weighted MRI also restricts the use of complementary information from other sequences or imaging planes. Although disc-level aggregation helps maintain anatomical consistency and reduce data leakage, it can limit more precise slice-level localization. Future work will therefore consider incorporating multi-sequence MRI, exploring ordinal or uncertainty-aware formulations for borderline cases, and improving spatial modeling to enhance localization and interpretability, with further validation on larger multi-institutional datasets.

\bibliographystyle{splncs04}  
\bibliography{references}      

\end{document}